\documentclass[showpacs,preprintnumbers,amsmath,amssymb,pra]{revtex4}
\usepackage{graphicx}
\usepackage{dcolumn}
\usepackage{bm}
\usepackage{times}
\usepackage[colorlinks,citecolor=blue,linkcolor=red]{hyperref}
\usepackage{color}
\usepackage{comment}

\begin{document}    

\title{Unifying quantum heat transfer and superradiant signature in a nonequilibrium collective-qubit system: a polaron-transformed Redfield approach}

\author{Xu-Min Chen$^{1}$}
\author{Chen Wang$^{2,}$}\email{wangchenyifang@gmail.com}
\address{
$^{1}$Department of Physics, Hangzhou Dianzi University, Hangzhou 310018, P. R. China \\
$^{2}$Department of Physics, Zhejiang Normal University, Jinhua 321004, Zhejiang , P. R. China
}

\date{\today}

\begin{abstract}
We investigate full counting statistics of quantum heat transfer in a collective-qubit system, constructed by
multi-qubits interacting with two thermal baths.
The nonequilibrium polaron-transformed Redfield approach embedded with an auxiliary counting field
is applied to obtain the steady state heat current and fluctuations,
which enables us to study the impact of the qubit-bath interaction in a wide regime.
The heat current, current noise and skewness are all found to clearly unify the limiting results
in the weak and strong couplings, respectively.
Moreover, the superradiant heat transfer is clarified as a system-size-dependent effect,
and large number of qubits dramatically suppresses the nonequilibrium superradiant signature.
\end{abstract}


\pacs{{05.60.Gg, 44.10.+i, 63.22.-m, 05.70.Ln} }


\maketitle

\section{Introduction}

Understanding mechanism of nonequilibrium quantum transport in low dimensional systems is a long-standing challenge,
which has been extensively investigated from solid-state physics~\cite{apjauho2008}, quantum thermodynamics~\cite{rkosloff2013entropy},
molecular electronics~\cite{anitzan2014} to quantum biology~\cite{mmohseni2014}.
In particular, quantum thermal transport  where the particle and heat flows are modulated by the temperature bias,
has triggered the emergence of the phononics~\cite{jswang2008epjb,nbli2012rmp,jren2015aip}.
{Phonons have been successfully utilized to fabricate various functional devices}, e.g., thermal diode, memory and transistor~\cite{bli2004prl,bli2006apl,lwang2007prl,lwang2008prl}.
In analogy with phononics, quantum heat transfer mediated by the spin(e.g., anharmonic molecule or qubit) has also been intensively analyzed
in a parallel way,
which leads to the realization of quantum spin-thermal transistor~\cite{kjoulain2016prl,cwang2018pra},
spin heat engine~\cite{dsegal2008prl,bsothmann2012epl,jren2013prb} and nonequilibrium spin network~\cite{gtcraven2016pnas,gtcraven2017prl}.

As a generic model to describe quantum heat transfer at the nanoscale,
the nonequilibrium spin-boson model(NESB) is composed by a two-level-system(TLS)
coupled to two thermal baths~\cite{dsegal2005prl},
which was originally proposed to study the quantum dissipation~\cite{ajleggett1987rmp,pao1989prl,uweiss2008}.
Many approaches have been explored to investigate the underlying mechanism of heat transfer
in the NESB~\cite{dsegal2006prb,mgalperin2007prb,kvelizhanin2008cpl,ksaito2013prl,yyao2015prb,etaylor2015prl,akato2015jcp,akato2016jcp,dzxu2016fp,cwang2017pra,jliu2017pre,lferialdi2017pra}.
{Analytically, the Redfield scheme is properly introduced to investigate the sequential process in the weak spin-bath interaction regime},
where two thermal baths show additive contribution to the heat transfer~\cite{dsegal2008prl,jren2010prl}.
{However, the Redfield approach breaks down in strong spin-bath interaction regime},
where the multiphonon excitations should be involved to characterize the nonequilibrium heat-exchange.
Then, the nonequilibrium noninteracting blip approximation(NIBA) can be applied to
study baths induced nonadditive and cooperative transfer processes~\cite{dsegal2006prb,lnicolin2011jcp,lnicolin2011prb,tchen2013prb,dsegal2014pre}.
However, both methods are found to have their own limitations, i.e.,
the Redfield approach is unable to capture the multi-phonon processes in strong coupling regime,
and the heat current based on the nonequilibrium NIBA scheme does not show linear proportion to coupling strength in weak coupling regime~\cite{tchen2013prb}.
Recently, the nonequilibrium polaron-transformed Redfield equation(NE-PTRE) was proposed to successfully
unify the steady state heat current in the NESB~\cite{cwang2017pra,cwang2015sr,dzxu2016njp}.
{
However, the exploration of the NE-PTRE to study more nonequilibrium spin systems is lacking, but is urgently required for the spin-based quantum heat transfer. This paper aims to fill this gap by applying the NE-PTRE to analyze the heat transfer in the collective-qubit model.}

Recently, a superradiant signature of quantum heat transfer has been discovered
in the collective-qubit system with weak qubit-bath interaction~\cite{mvogl2011aop,mvogl2012pra}.
The steady state superradiance describes the effect that qubits collectively exchange energy with thermal baths,
resulting in the current scaling as $J{\sim}N^2_s$ with $N_s$ the number of  qubits.
In sharp contrast, the superradiant heat transfer {vanishes} in strong coupling regime, based on the nonequilibrium NIBA under the Marcus approximation~\cite{cwang2015aop}.
Hence, the steady state behavior of the collective-qubit system are significant distinct from each other in limiting interaction regimes.
It is demanding to analyze the heat transfer feature from weak to strong couplings from a unified perspective。

To address these problems, we extend the NE-PTRE combined with full counting statistics(FCS)~\cite{mesposito2009rmp,mcampisi2011rmp} to investigate quantum heat transfer in the collective-qubit system.
The counting-field dependent NE-PTRE successfully unifies  the current and fluctuations(e.g., heat current, current noise, skewness), with the incoherent picture in the weak quit-bath coupling limit
and multi-phonon excited transfer picture in the strong coupling limit, respectively.
Moreover, the superradiant heat transfer is investigated at large temperature bias~\cite{mvogl2011aop},
and the disappearance of the superradiant signature is explained by
enlarging the number of qubits beyond the weak qubit-bath interaction.

This paper is organized as follows:
In Sec. II, the collective-qubit model and the NE-PTRE are described.
In Sec. III, FCS is briefly introduced and
the counting-field dependent NE-PTRE is derived, which enables us
to analyze the steady state heat transfer.
In Sec. IV, heat current, current noise and skewness are all found to hold the unified features,
by extending the application of the NE-PTRE.
In Sec. V, we study the transition of superradiant heat transfer from weak to strong couplings,
and explain the vanishing mechanism of superradiant signature with large number of qubits.
In the final section, we present a concise summary.

\section{Nonequilibrium collective-qubit system}

\subsection{Model}
The nonequilibrium energy transfer in the collective-qubit system, which interacts with two thermal baths, is modeled as
$\hat{H}=\hat{H}_s+\sum_{u=L,R}(\hat{H}^u_b+\hat{V}_u)$.
The collective-qubit model is described as
\begin{eqnarray}~\label{ham1}
\hat{H}_s=\varepsilon\hat{J}_z+\Delta\hat{J}_x,
\end{eqnarray}
where the collective angular-momentum operators are
$\hat{J}_{a}=\frac{1}{2}\sum^{N_s}_{i}\hat{\sigma}^i_a~(a=x,y,z)$ with $\hat{\sigma}^i_a$ the Pauli operator of the $i$th qubit, $N_s$ is the number of qubits,
$\varepsilon$ and $\Delta$ are the Zeeman splitting energy and coherent tunneling strength of the angular momentum operator.
In the limit of $N_s=1$, the Hamiltonian at Eq.~(\ref{ham1}) is reduced to the seminal nonequilibrium spin-boson model~\cite{dsegal2005prl,jren2010prl}.
The $u$th thermal bath is composed by noninteracting bosons, shown as
$\hat{H}^u_b=\sum_k\omega_k\hat{b}^{\dag}_{k,u}\hat{b}_{k,u}$,
where $\hat{b}^{\dag}_{k,u}(\hat{b}_{k,u})$ creates (annihilates) one phonon with the frequency $\omega_k$.
The system-bath interaction is given by
\begin{eqnarray}
\hat{V}_u=2\hat{J}_z\sum_k(g_{k,u}\hat{b}^{\dag}_{k,u}+g^*_{k,u}\hat{b}_{k,u}),
\end{eqnarray}
where $g_{k,u}$ is the coupling strength between the angular-momentum and the $u$th thermal bath.
{It is characterized by the spectral function $G_u(\omega)=4\pi\sum_k|g_{k,u}|^2\delta(\omega-\omega_k)$.}
In this paper, we select the spectral function
as the super-Ohmic form $G_u(\omega)=\pi\alpha_u\omega^3/\omega^2_c\exp(-\omega/\omega_c)$,
where $\alpha_u$ is the coupling strength and $\omega_c$ is the cutoff frequency of thermal baths.
The super-Ohmic spectrum has been extensively considered to investigate the
quantum dissipative dynamics and transport in molecular electronics~\cite{anazir2009prl,sjjang2013njp,cklee2015jcp},
solid-state devices~\cite{dpsm2013prl} and light-harvesting complexes~\cite{sjjang2011jcp,dpsm2011prb}.

To analyze the multi-phonon involved energy transfer processes, we apply the canonical transformation
$\hat{U}=\exp[i\hat{J}_z\sum_{u}\hat{B}_u]$ to the Hamiltonian $\hat{H}$ as
$\hat{H}^{\prime}=\hat{U}^{\dag}\hat{H}\hat{U}=\hat{H}^{\prime}_s+\sum_{u=L,R}(\hat{H}^u_b+\hat{V}^{\prime}_u)$~\cite{rsilbey1984jcp,rharris1985jcp},
where the collective phonon momentum is $\hat{B}_u=2i\sum_k(\frac{g_{k,u}}{\omega_k}\hat{b}^{\dag}_{k,u}
-H.c.)$.
After the transformation, the modified system Hamiltonian is given by
\begin{eqnarray}~\label{hp1}
\hat{H}^{\prime}_s=\varepsilon\hat{J}_z+\eta\Delta\hat{J}_x-{\xi}\hat{J}^2_z,
\end{eqnarray}
where the factor $\eta={\langle}\cos{\hat{B}}{\rangle}$ is specified as
\begin{eqnarray}
\eta=\exp{[-2\sum_{k,u=L,R}|g_{k,u}/\omega_k|^2(2n_{k,u}+1)]},
\end{eqnarray}
and the renormalization energy is $\xi=4\sum_{k,u}|g_{k,u}|^2/\omega_k$.
In weak system-bath interaction limit $|g_{k,u}/\omega_k|^2{\approx}0$, the renormalization factor becomes $\eta=1$
and $\xi{\approx}0$.
While in  strong system-bath interaction region $|g_{k,u}/\omega_k|^2{\gg}1$,
the factor $\eta$ approaches zero, and the renormalization energy $\xi{\gg}\{\varepsilon,\Delta\}$.
{The generation of large $\xi$ will dramatically change the energy structure of the modified qubits system, compared to the counterpart in the weak coupling limit.}
Moreover, the modified system-bath interaction is given by
\begin{eqnarray}
\hat{V}^{\prime}_u={\Delta}[(\cos{\hat{B}}-\eta)\hat{J_x}+\sin{\hat{B}}\hat{J}_y].
\end{eqnarray}

\subsection{Nonequilibrium polaron-transformed Redfield equation}

We apply the NE-PTRE to investigate the dynamics of the collective-qubit model.
The NE-PTRE, one type of the quantum master equation, has bee successfully applied to
unify the nonequilibrium energy transfer in the seminal spin-boson model~\cite{cwang2017pra,dzxu2016njp}.
It is known that the modified system-bath interaction disappears under the thermal average (${\langle}\hat{V}^{\prime}_u{\rangle}$=0).
Hence, it may be safe to perturb ${\langle}\hat{V}^{\prime}_u{\rangle}$ up to the second order to obtain the quantum master equation.
Under the Born approximation, the density operator of the whole system can be decomposed by
$\hat{\rho}(t)=\hat{\rho}_s(t){\otimes}\hat{\rho}_b$ {with $\hat{\rho}_s(t)$ the density operator of the qubits and
$\hat{\rho}_b=e^{-\sum_u\hat{H}^u_b/k_BT_u}/\textrm{Tr}_b\{e^{-\sum_u\hat{H}^u_b/k_BT_u}\}$ the density operator of  the baths at equilibrium.}
The quantum master equation in the Markovian limit is obtained as
\begin{eqnarray}
\frac{d\hat{\rho}_s(t)}{dt}&=&-i[\hat{H}^{\prime}_s,\hat{\rho}_s(t)]
+\sum_{a=x,y}\int^\infty_0d{\tau}{\times}\nonumber\\
&&(C_a(\tau)[\hat{J}_a(-\tau)\hat{\rho}_s(t),\hat{J}_a]+H.c.),
\end{eqnarray}
where the correlation functions are
\begin{eqnarray}~\label{corr1}
C_x(\tau)&=&(\eta\Delta)^2[{\cosh}(\sum_uQ_u(\tau))-1],\\
C_y(\tau)&=&(\eta\Delta)^2{\sinh}(\sum_uQ_u(\tau)),\nonumber
\end{eqnarray}
with the phonon propagator in $u$th thermal bath
$Q_u(\tau)=4\sum_k|g_{k,u}/\omega_k|^2[\cos\omega_k\tau(2n_{k,u}+1)-i\sin\omega_k\tau]$.
Furthermore, in the eigen-basis $\hat{H}^{\prime}_s|\phi_n{\rangle}=E_n|\phi_n{\rangle}$, the dynamical equation can be re-expressed as
\begin{eqnarray}~\label{qme1}
\frac{d{\rho}_{nn^{\prime}}}{dt}&=&-iE_{nn^{\prime}}{\rho}_{nn^{\prime}}
+\sum_{a=x,y}[(\Gamma_a(E_{nm})+\Gamma^*_a(E_{n^{\prime}m^{\prime}}))J^{nm}_aJ^{m^{\prime}n^{\prime}}_a{\rho}_{mm^{\prime}}\nonumber\\
&&-\Gamma_a(E_{mm^{\prime}})J^{nm}_aJ^{mm^{\prime}}_a{\rho}_{m^{\prime}n^{\prime}}
-\Gamma^*_a(E_{m^{\prime}m})J^{mm^{\prime}}_aJ^{m^{\prime}n^{\prime}}_a{\rho}_{nm}]
\end{eqnarray}
where the transition rate is
$\Gamma_a(\omega)=\int^\infty_0{d\tau}C_a(\tau)e^{-i\omega\tau}$ {and the element is ${\rho}_{nn^{\prime}}={\langle}\phi_n|\hat{\rho}_s(t)|\phi_{n^{\prime}}{\rangle}$.}
The rate $\Gamma_{x(y)}(E_{nm})$ describes that even(odd) number phonons are involved in the transfer process between the states $|\phi_n{\rangle}$
and $|\phi_m{\rangle}$.

In the weak qubit-bath coupling limit, the heat transfer is dominated by the sequential process and $\eta{\approx}1$.
{Thus, the correlation function $C_y(\tau)$ is reduced to $C_y(\tau){\approx}\Delta^2[\sum_uQ_u(\tau)]$,
and $C_x(\tau){\approx}0$ by ignoring high-order correlations.}
Accordingly, the lowest order of the transition rate $\Gamma^{(1)}_y(\omega)$ includes the term
$\Delta^2(Q_L(\omega)+Q_R(\omega))$, with $Q_v(\omega)=\int{d\tau}e^{-i\omega\tau}Q_v(\tau)$.
Moreover, all of off-diagonal elements of the qubits density matrix approach zero at steady state
(not shown here, which is quite similar to the result at Fig.~\ref{fig-app-a}).
Considering the commutating relation {$[\hat{H}^{\prime}_s,\hat{J}_z]=-i{\Delta}\hat{J}_y$},
the quantum master equation at Eq.~(\ref{qme1}) after long time evolution is simplified as
\begin{eqnarray}~\label{qme2}
\frac{dP_{nn}}{dt}&=&2\sum_{m{\neq}n,u}{G_u(E_{nm})}n_u(E_{nm})J^{nm}_{z}J^{mn}_{z}\rho_{mm}\\
&&-2\sum_{m{\neq}n,u}{G_u(E_{mn})}n_u(E_{mn})J^{nm}_{z}J^{mn}_{z}\rho_{nn}\nonumber,
\end{eqnarray}
which is identical with the dynamical equation based on the Redfield scheme (see appendix A for the detail).

While in strong qubit-bath interaction limit, the coherent tunneling of qubits at Eq.~(\ref{hp1}) is dramatically suppressed ($\eta{\approx}0$), and the correlation factor $\eta^2e^{-\sum_uQ_u(\tau)}$ vanishes.
However, the other factor $\eta^2e^{\sum_uQ_u(\tau)}$ is kept finite, which contributes to the quantum heat transfer.
Hence, the correlation functions at Eq.~(\ref{corr1}) are reduced to
$C_x(\tau)=C_y(\tau)=(\eta\Delta)^2[\exp{\sum_uQ_u(\tau)}]$.
Consequently, the master equation at Eq.~(\ref{qme1}) in the local basis $\{|\phi_n{\rangle}\}$ with
$(\varepsilon\hat{J}_z-\xi\hat{J}^2_z)|\phi_n{\rangle}=(\varepsilon{n}-\xi{n}^2)|\phi_n{\rangle}$ is changed to
\begin{eqnarray}~\label{niba1}
\frac{dP_n}{dt}=-(\kappa^-_{n-1}+\kappa^+_n)P_n+\kappa^-_{n}P_{n+1}
+\kappa^+_{n-1}P_{n-1},
\end{eqnarray}
where the rate is
$\kappa^{\pm}_n=\frac{(j^+_n\Delta)^2}{8\pi}\int{d\omega}C_L(\pm\omega\mp\Delta_n)C_R(\mp\omega)$,
with $j^+_n=\sqrt{\frac{N_s}{2}(\frac{N_s}{2}+1)-n(n+1)}$, the energy gap $\Delta_n=\varepsilon-\xi(2n+1)$
and the correlation function in the frequency domain
$C_v(\omega)=\int{dt}e^{i\omega{t}}\eta^2_ve^{Q_v(t)}$.
Hence, the dynamical equation is completely reduced to the nonequilibrium NIBA scheme (see Eq.(\ref{qme-niba}) at appendix B for the detail).

\section{Full counting statistics of heat current}

\subsection{The general theory}
Full counting statistics is a two-time projection protocol to measure the current and fluctuations~\cite{mesposito2009rmp,mcampisi2011rmp}.
For the energy transfer in the multi-terminal setup, the generating function is generally given by~\cite{hmf2018njp}
\begin{eqnarray}
Z(\{\chi_u\},t)=\textrm{Tr}[e^{i\sum_u\chi_u\hat{H}_u(0)}e^{-i\sum_u\chi_u\hat{H}_u(t)}\hat{\rho}_{tot}(0)],
\end{eqnarray}
where $\chi_u$ is the counting-field parameter to count the energy flow into the $u$th bath with the Hamiltonian $\hat{H}_u$,
$\hat{H}_u(t)=\hat{U}^{\dag}\hat{H}_u\hat{U}$ with the propagator $\hat{U}=e^{-i\hat{H}t}$,
and $\hat{\rho}_{tot}(0)$ is the initial density matrix of the whole system.
Moreover, considering the modified propagator
$\hat{U}_{\{\chi_u\}}(t)=e^{i\sum_u\chi_u\hat{H}_u/2}\hat{U}(t)e^{-i\sum_u\chi_u\hat{H}_u/2}$,
it can be re-expressed as
$\hat{U}_{\{\chi_u\}}(t)=\exp{(-i\hat{H}_{\{\chi_u\}}t)}$ with
$\hat{H}_{\{\chi_u\}}=e^{i\sum_u\chi_u\hat{H}_u/2}\hat{H}e^{-i\sum_u\chi_u\hat{H}_u/2}$.
Hence, the generating function is re-expressed as
\begin{eqnarray}
Z(\{\chi_u\},t)=\textrm{Tr}[\hat{U}_{-\{\chi_u\}}\hat{\rho}_{tot}(0)\hat{U}^{\dag}_{\{\chi_u\}}]=\textrm{Tr}[\hat{\rho}^{\{\chi_u\}}_{tot}(t)].
\end{eqnarray}
After the long-time evolution, the cumulant generating function is obtained as
\begin{eqnarray}
\Theta(\{\chi_u\})=\lim_{t\rightarrow\infty}\frac{1}{t}\ln{Z(\{\chi_u\},t)}.
\end{eqnarray}
Therefore, the $n$th cumulant of heat current fluctuations into the $u$th bath is given by
\begin{eqnarray}~\label{current1}
J^{(n)}_u={{\frac{{\partial}^n\Theta(\{\chi_u\})}{{\partial}(i\chi_u)^n}}}\bigg{|}_{\{\chi_u=0\}}.
\end{eqnarray}
Specifically, the lowest cumulant is the steady state heat current
$J_u={\frac{{\partial}\Theta(\{\chi_u\})}{{\partial}(i\chi_u)}}\Big{|}_{\{\chi_u=0\}}$,
the second cumulant is the zero-frequency current noise
$J^{(2)}_u={\frac{{\partial}^2\Theta(\{\chi_u\})}{{\partial}(i\chi_u)^2}}\Big{|}_{\{\chi_u=0\}}$
and the third cumulant is the current skewness
$J^{(3)}_u={\frac{{\partial}^3\Theta(\{\chi_u\})}{{\partial}(i\chi_u)^3}}\Big{|}_{\{\chi_u=0\}}$.
In the following, we apply the theory of the FCS to obtain the
counting-field dependent NE-PTRE of the nonequilibrium collective-qubit model.

\subsection{Counting-field dependent NE-PTRE}
To count the heat flow into the right thermal bath, we introduce the counting field parameters as
$\hat{H}_\chi=e^{i\hat{H}^R_b\chi/2}\hat{H}e^{-i\hat{H}^R_b\chi/2}$~\cite{hmf2018njp}, which results in
\begin{eqnarray}~\label{hchi}
\hat{H}_\chi=\hat{H}_s+\sum_{u=L,R}(\hat{H}^u_b+\hat{V}_u(\chi)),
\end{eqnarray}
where the system-bath interaction with the counting field parameter is expressed as
\begin{eqnarray}
\hat{V}_u(\chi)=2\hat{J}_z\sum_k(g_{k,u}e^{i\frac{\omega_k\chi}{2}\delta_{u,R}}\hat{b}^{\dag}_{k,u}
+H.c.),
\end{eqnarray}
{where $\delta_{R,R}=1$ and $\delta_{L,R}=0$}.
Then, after the canonical transformation $\hat{H}^{\prime}_{\chi}=U^{\dag}_{\chi}\hat{H}_\chi{U_\chi}$
with transformation operator $U=\exp[i\hat{J}_z\sum_u\hat{B}_u(\chi)]$ and
$\hat{B}_u(\chi)=2i\sum_k(\frac{g_{k,u}}{\omega_k}e^{i\frac{\omega_k\chi}{2}\delta_{u,R}}\hat{b}^{\dag}_{k,u}-H.c.)$,
the modified Hamiltonian is given by
\begin{eqnarray}
\hat{H}^{\prime}_{\chi}=\hat{H}^{\prime}_s+\sum_{u=L,R}(\hat{H}^u_b+\hat{V}^{\prime}_u(\chi)),
\end{eqnarray}
where the modified system-bath interaction with counting field parameter is given by
\begin{eqnarray}
\hat{V}^{\prime}_u(\chi)=\Delta[(\cos\hat{B}_{\chi}-\eta)\hat{J}_x+\sin\hat{B}_{\chi}\hat{J}_y].
\end{eqnarray}

{By perturbing $\hat{V}^{\prime}_u(\chi)$ under the Born-Markov approximation}, we obtain the second-order quantum master equation as
\begin{eqnarray}~\label{qme-chi-2}
\frac{d\hat{\rho}_{\chi}}{dt}&=&-i[\hat{H}^{\prime}_s,\hat{\rho}_{\chi}]
+\sum_a\int^\infty_0d{\tau}[C_a(\chi,-\tau)\hat{J}_a\hat{\rho}_{\chi}\hat{J}_a(-\tau)\nonumber\\
&&+C_a(\chi,\tau)\hat{J}_a(-\tau)\hat{\rho}_{\chi}\hat{J}_a-C_a(\tau)\hat{J}_a\hat{J}_a(-\tau)\hat{\rho}_{\chi}\nonumber\\
&&-C_a(-\tau)\hat{\rho}_{\chi}\hat{J}_a(-\tau)\hat{J}_a],
\end{eqnarray}
where the correlation functions with counting field parameter are
\begin{eqnarray}
C_x(\chi,\pm\tau)&=&(\eta\Delta)^2[\cosh(Q_L(\pm\tau)+Q_R(\pm\tau-\chi))-1],\nonumber\\
C_y(\chi,\pm\tau)&=&(\eta\Delta)^2\sinh(Q_L(\pm\tau)+Q_R(\pm\tau-\chi)).
\end{eqnarray}
In absence of the counting field parameter, the correlation functions are reduced to the standard version at Eq.~(\ref{corr1}).
In the eigen-basis, the dynamics of density matrix elements can be specified as
\begin{eqnarray}~\label{qmechi1}
\frac{d\rho^{\chi}_{nn^{\prime}}}{dt}&=&-iE_{nn^{\prime}}\rho^{\chi}_{nn^{\prime}}
+\sum_a[(\Gamma^{\chi}_{a,-}(E_{m^{\prime}n^{\prime}})+\Gamma^{\chi}_{a,+}(E_{nm}))J^{nm}_aJ^{m^{\prime}n^{\prime}}_a\rho^{\chi}_{mm^{\prime}}\nonumber\\
&&-\Gamma_a(E_{mm^{\prime}})J^{nm}_aJ^{mm^{\prime}}_a\rho^{\chi}_{m^{\prime}n^{\prime}}
-\Gamma^*_a(E_{m^{\prime}m})J^{mm^{\prime}}_aJ^{m^{\prime}n^{\prime}}_a\rho^{\chi}_{nm}]
\end{eqnarray}
where the transition rates are
\begin{eqnarray}
\Gamma^{\chi}_{a,\pm}(\omega)=\int^\infty_0d{\tau}C_a(\chi,\pm\tau)e^{-i\omega\tau}.
\end{eqnarray}

If we reorganize the reduced density matrix of collective-qubit from the matrix form to the vector expression,
the dynamical equation at Eq.~(\ref{qmechi1}) is expressed as
\begin{eqnarray}
\frac{d}{dt}|\rho_{\chi}(t){\rangle}=\hat{\mathcal{L}}(\chi)|\rho_{\chi}(t){\rangle},
\end{eqnarray}
where $\hat{\mathcal{L}}(\chi)$ is the superoperator to dominate the system evolution.
Then, the cumulant generating function at $t$-time is given by
\begin{eqnarray}
\Theta(\chi,t)=\frac{1}{t}\ln[{\langle}I|e^{\hat{\mathcal{L}}(\chi)t}|\rho(0){\rangle}],
\end{eqnarray}
where $|\rho(0){\rangle}$ is the vector form of the initial system density matrix,
and ${\langle}I|$ is the left-eigenvector of $\hat{\mathcal{L}}$ as ${\langle}I|\hat{\mathcal{L}}=0$,
with the normalization relation ${\langle}I|\rho_{\chi=0}(t){\rangle}=1$.
Hence, heat current fluctuations at steady state can be straightforwardly obtained by following the Eq.~(\ref{current1}).

\begin{figure*}[tbp]
\begin{center}
\vspace{-1.0cm}
\includegraphics[scale=0.6]{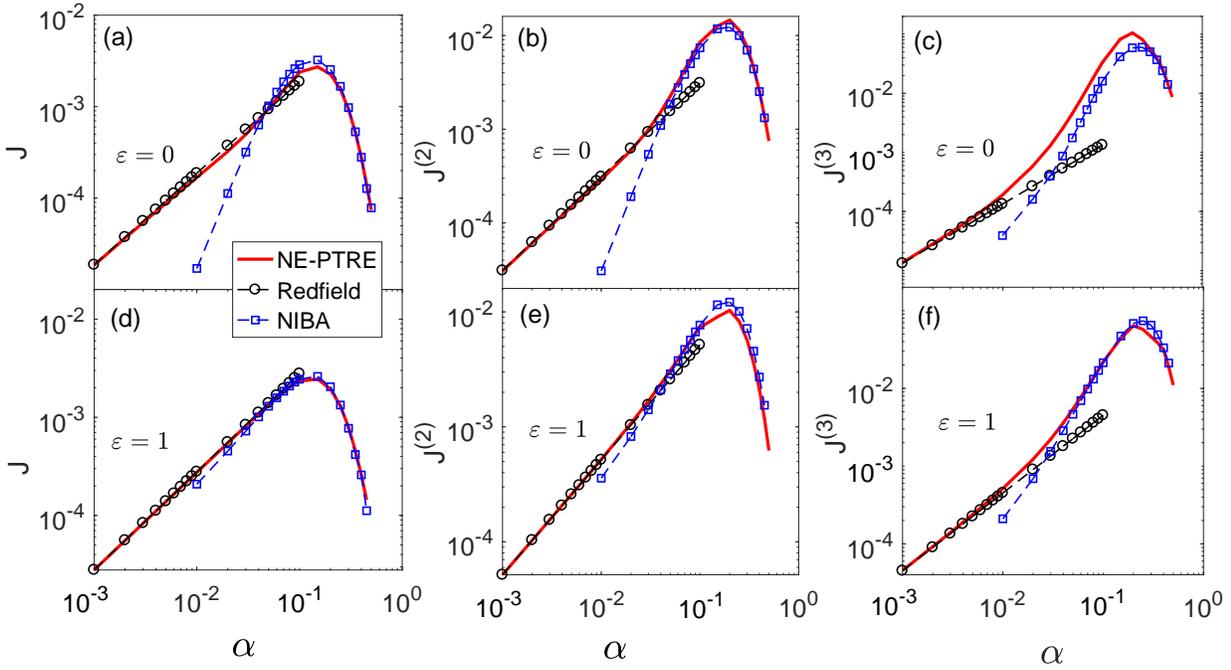}
\end{center}
\caption{(Color online) Comparisons of steady state current fluctuations based on the NE-PTRE with counterparts within the Redfield and NIBA schemes:
(a),(b),(c) at resonance ($\varepsilon=0$) and (d),(e),(f) at off-resonance ($\varepsilon=1$),
by tuning qubit-bath coupling strength $\alpha$.
The other parameters are given by $N_s=2$, $\Delta=1$, $\omega_c=6$, $T_L=1.5$ and $T_R=0.5$.
}~\label{fig1}
\end{figure*}

\begin{figure*}[tbp]
\includegraphics[scale=0.55]{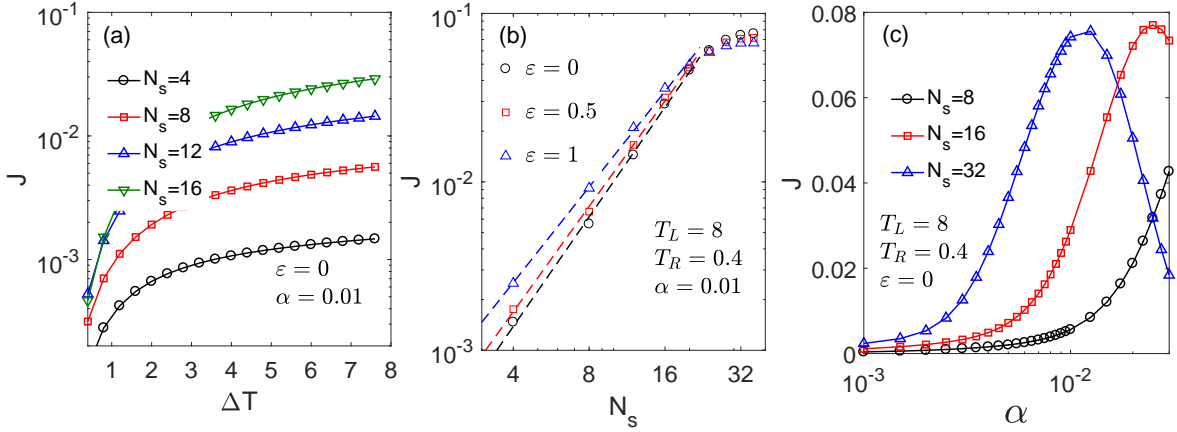}
\vspace{-1.0cm}
\caption{(Color online) Behaviors of steady state heat current $J$ by tuning
(a) temperature bias $\Delta{T}=T_L-T_R$ with $T_R=0.4$, $\varepsilon=0$ and $\alpha=0.01$;
(b) the number of qubits $N_s$ with $T_L=8$, $T_R=0.4$ and $\alpha=0.01$;
(c) qubit-bath coupling strength $\alpha$ with $T_L=8$, $T_R=0.4$ and $\varepsilon=0$.
The other parameters are $\Delta=1$ and $\omega_c=6$.
}~\label{fig2}
\end{figure*}

\section{Unified steady state heat current}

Quantum heat transfer in the NESB has been successfully investigated based on the Redfield and noninteracting-blip approximation schemes in the weak and strong qubit-bath coupling limits, separately.
However, steady state heat current was found to be distinct from each other in a broad coupling regime~\cite{tchen2013prb}.
Recently, the nonequilibrium polaron-transformed Redfield equation combined with full counting statistics was proposed
to unify heat current between these two limiting approaches~\cite{cwang2017pra,dzxu2016njp}.

Here, we try to extend the counting-field dependent NE-PTRE to unify the heat transfer in the nonequilibrium collective-qubit model  in Fig.~\ref{fig1}.
At resonance($\varepsilon=0$), we first analyze the steady state heat current in Figs.~\ref{fig1}(a) and \ref{fig1}(d).
It clearly exhibits the turnover behavior, which unifies the counterparts within the Redfield and NIBA methods as the qubit-bath coupling strength
approaches the weak and strong coupling limits.
We admit that to gain an analytical expression of heat current at arbitrary qubit-bath coupling is rather difficult.
However, it can be obtained in limiting regimes.
{
Here, we study the analytical expression of the steady state heat current with $N_s=2$.
Specifically, in the weak interaction limit the heat current is analytically expressed as(see Eq.~(\ref{app-a-jweak}) at appendix A)
\begin{eqnarray}~\label{jweak}
J_{\textrm{weak}}=4[n_L(\Delta)-n_R(\Delta)]\frac{G_L(\Delta)G_R(\Delta)}{G_L(\Delta)+G_R(\Delta)}\frac{(2-x)(1+x^3)}{(1+x+x^2)},
\end{eqnarray}
where the coefficient $x=[\sum_{v=L,R}G_v(\Delta)n_v(\Delta)]/[\sum_{v=L,R}G_v(\Delta)(1+n_v(\Delta))]$,
which is linearly proportional to the qubit-bath coupling strength.
The current at Eq.~(\ref{jweak}) is the special case of the general expression at Eq.~(\ref{app-a-jweak}).}
It is found in Fig.\ref{fig1}(a) that $J_{\textrm{weak}}$ becomes identical with the counterpart from NE-PTRE in weak coupling limit (e.g., $\alpha=0.001$).
{Moreover, it should be noted that the current $J_{\textrm{weak}}$ at Eq.~(\ref{jweak}) with $N_s=2$ has the similar expression to the case in the standard NESB($N_s=1$) in the weak coupling limit, which both proportional to the thermodynamic bias (i.e., $J{\propto}[n_L(\Delta)-n_R(\Delta)]$)~\cite{dsegal2006prb}.}

While in the strong coupling limit, the dynamical equation at Eq.~(\ref{qme-chi-2}) with the number of qubits $N_s=2$ is reduced to the kinetic form as (see Eq.~(\ref{qme-niba}) at appendix B)
\begin{eqnarray}~\label{qme-niba}
\frac{dP^\chi_n(t)}{dt}&=&-(\kappa^-_{n-1}+\kappa^+_n)P^\chi_n(t)
+\kappa^-_{n}(\chi)P^\chi_{n+1}(t)+\kappa^+_{n-1}(\chi)P^\chi_{n-1}(t)~(n=-1,0,1),
\end{eqnarray}
where the population is $P^{\chi}_n={\langle}1,n|\hat{\rho}_{\chi}(t)|1,n{\rangle}$ with $\hat{J}_z|1,n{\rangle}=n|1,n{\rangle}$.
The transition rate is
\begin{eqnarray}
\kappa^{\pm}_n(\chi)&=&\frac{(j^+_{n}\Delta)^2}{8\pi}\int^\infty_{-\infty}d\omega{e^{\mp{i\omega\chi}}}
C_R(\mp\omega)C_L(\pm\omega\mp\Delta_n),\nonumber\\
\end{eqnarray}
where the coefficient is $j^+_n=\sqrt{2-n(n{+}1)}$,
the energy gap $\Delta_n=E_{n+1}-E_n={\varepsilon}-\xi(2n+1)$
and the correlation function in the frequency domain is
$C_v(\omega)=\int^\infty_{-\infty}dte^{i\omega{t}}\eta^2_ve^{Q_v(t)}$
with the factor $\eta_v=\exp{[-2\sum_{k}|g_{k,v}/\omega_k|^2(2n_{k,v}+1)]}$.
{If we select $N_s=1$($j^+_{-1/2}=1$), the rate $\kappa^{\pm}_n(\chi)$ is reduced to the standard NESB case(see Eq.~(20) in Ref.~\cite{lnicolin2011jcp}).}
Thus, the cumulant generating function at steady state is given by
\begin{eqnarray}
\Theta_{\textrm{NIBA}}(\chi)&=&\sqrt{(2\kappa^+_0-\kappa^-_0)^2+8\kappa^+_0(\chi)\kappa^-_0(\chi)}\Big{/}2\nonumber\\
&&-(2\kappa^+_0+\kappa^-_0)/2,
\end{eqnarray}
where $\kappa^{\pm}_{-1}(\chi)=\kappa^{\mp}_{0}(\chi)$.
Consequently, the heat current is given by
\begin{eqnarray}~\label{jstrong}
J_{\textrm{strong}}=\frac{\Delta^2}{2\pi}\int^{\infty}_{-\infty}d\omega\omega
\left[\frac{\kappa^-_0}{2\kappa^+_0+\kappa^-_0}C_R(\omega)C_L(-\omega+\xi)
-\frac{\kappa^+_0}{2\kappa^+_0+\kappa^-_0}C_R(-\omega)C_L(\omega-\xi)\right],
\end{eqnarray}
where the first(second) term describes the process that as the qubits release(gain) energy $\xi$, the right bath absorbs(emits) phonon energy $\omega$
and the left bath obtains(provides) the remaining energy $\xi-\omega$.
{$J_{\textrm{strong}}$ shows similar structure with the counterpart in the standard NESB, which is jointly contributed by two thermal baths~\cite{lnicolin2011jcp}}.
Moreover, the expression of $J_{\textrm{strong}}$ captures the turnover feature of heat current in Fig.~\ref{fig1}(a) and \ref{fig1}(d) with dashed blue lines with squares.
Hence, we conclude that the steady state heat flux is clearly unified in the nonequilibrium collective-qubit model.

Next, we analyze the zero-frequency current noise and skewness in Figs.~\ref{fig1}(b-c) and \ref{fig1}(e-f).
It is interesting to find that the results based on the NE-PTRE also perfectly  bridge the limiting counterpart at the weak and strong coupling regimes,
which may demonstrate the unification of current fluctuations in extended spin-boson systems (e.g., collective-qubit model).
{Moreover, the turnover behavior which is presented in the current, are unraveled for the second and third cumulants.}
Though not shown here, higher cumulants of current fluctuations also show such unified features.
{We should note that all above results are valid both at resonant and off-resonant conditions,
which clearly exhibits that full counting statistics of the heat current is generally unified within the NE-PTRE scheme.}

\section{Suppression of superradiant heat transfer}

Superradiant effect, which describes that the system exhibits collective response under the modulation of the external field or thermal bath,
has been extensively investigated in quantum phase transition~\cite{nlambert2004prl,qhchen2008pra},
critical heat engine~\cite{auch2015sr} and quantum transport~\cite{mvogl2011aop,mvogl2012pra}.
In particular, quantum heat transfer in the nonequilibrium large-spin system shows the superradiant signature in the weak spin-bath coupling regime~\cite{mvogl2011aop}.
Specifically,  under large temperature bias (e.g., $T_L{\gg}T_R$) the steady state heat current is expressed as
\begin{eqnarray}~\label{curr-sr}
J=\frac{(1-x)}{3x}(n_L(\Delta)-n_R(\Delta))\frac{G_L(\Delta)G_R(\Delta)}{G_L(\Delta)+G_R(\Delta)}N^2_s,
\end{eqnarray}
which follows the condition $x/(1-x){\gg}N_s$.
{It should be noted that Eq.~(\ref{curr-sr}) is a special case of the current at Eq.~(\ref{app-a-jweak}) at appendix A.}
However, with the strong spin-bath interaction based on the NIBA scheme combined with the Marcus approximation, it is found that such superradiant transfer vanishes~\cite{cwang2015aop}.
Hence, we apply the NE-PTRE to clarify such seemingly paradox.

We first study the effect of the temperature bias $\Delta{T}$ on the heat current in Fig.~\ref{fig2}(a)
with spin-bath coupling strength $\alpha=0.01{\ll}\{\Delta,\omega_c\}$, which is considered weak for the seminal spin-boson model($\eta{\approx}1$).
It is found that the current shows monotonic enhancement by increasing both ${\Delta}T$ and $N_s$.
Moreover, at large temperature bias (e.g., $T_L=8$ and $T_R=0.4$) the current becomes nearly stable with large qubits number,
which is clearly exhibited in Fig.~\ref{fig2}(b) (e.g., $N_s=32$).
While for relatively small number of qubits (e.g., $N_s{<}20$) in Fig.~\ref{fig2}(b),
the superradiant signature of heat current is numerically obtained as
$J{\propto}N^{\gamma}_s$ with $\gamma=2.0{\pm}0.1$ (both for resonance and off-resonance).
Thus, it is known that the expression of superradiant heat current at Eq.~(\ref{curr-sr}) becomes invalid at large $N_s$.

{
To exploit the origin, we plot the current as a function of the qubit-bath coupling strength $\alpha$ for different $N_s$
with $T_L=8$, $T_R=0.4$ and $\varepsilon=0$, as shown in Fig.~\ref{fig2}(c).
It is shown that for small $N_s$(e.g., $N_s=8$), the heat flux exhibits approximately linear increasing behavior
in the weak coupling regime (e.g., $\alpha{\le}0.01$).
The heat flux demonstrates the sequential transfer process, where a superradiant heat transfer is observed accordingly,
which could be described by the Redfield scheme.
While the current shows distinct behavior for large $N_s$ from that for small $N_s$ in the same coupling regime.
For example, the current for $N_s=32$ with the coupling strength $\alpha=0.01$ has already surpassed the turnover point of the current,
while the current for $N_s=8$ is almost linearly dependent on the coupling strength in same coupling regime.
It is known that the appearance of the turnover point of the current is the significant signature of the multi-phonons involved coherent transfer,
as exploited in the nonequilibrium spin-boson model~\cite{cwang2017pra}.
Since $\eta$ is nearly equal to $1$ in above two cases and the heat current shows different features in same coupling regime, we conclude that $\eta{\approx}1$ does not necessarily correspond to the weak coupling condition.
Furthermore, the behavior of the current should be properly described by the NIBA scheme.
It results in the absence of nonequilibrium superradiant signature.
Therefore, the effect of superradiant heat transfer in the collective-qubit model will be dramatically suppressed in the large $N_s$ regime,
i.e. it is a $N_s$-dependent phenomenon. }


\section{Conclusion}
To summarize, we investigate the quantum heat transfer in the nonequilibrium collective-qubit system,
by applying the nonequilibrium polaron-transformed Redfield equation combined with full counting statistics.
We first analyze the effect of qubit-bath coupling on the steady state heat current,
which resulting in a turnover behavior, which is similar to the counterpart in the nonequilibrium spin-boson model.
Interestingly, the current consistently bridges the results in the weak and strong coupling limits,
which clearly demonstrates that heat current can be unified in the multi-qubits case.
It should be admitted that the general solution of the heat current is difficult to obtain even for $N_s=2$.
However, the analytical expression can be still obtained
in the weak and strong couplings based on the Redfield(Eq.~(\ref{jweak})) and NIBA(Eq.~(\ref{jstrong})) schemes, respectively.
Moreover, the current noise and skewness are shown to be unified accordingly.
Though not shown in the present paper, the unification of higher cumulants of heat current can also be observed.
Therefore, we propose that full counting statistics of heat current at steady state can be unified by using the NE-PTRE.

Next, we study the superradiant heat transfer at high temperature bias regime.
It is found that with small number of qubits, the heat transfer is described by the sequential process under the Redfield scheme.
The heat current exhibits $J{\sim}N^{2.0{\pm}0.1}_s$, an apparent signature of the steady state superradiance.
While with the large number of qubits, the superradiant signature of the heat flux vanishes.
{The corresponding physical process is described by the NIBA scheme, for multi-phonons should be involved to contribute to the heat transfer.}
Therefore, we conclude that the superradiant transfer in the  collective-qubit model is a size-dependent phenomenon,
and it will be strongly suppressed by increasing the qubits number.

We believe that based on the counting-field dependent NE-PTRE, the unified feature of steady state heat transfer
may be realized in a much bigger family of the quantum nonequilibrium system, e.g., quantum spin-boson network~\cite{gtcraven2017prl}.

\section{Acknowledgement}
X.M.C acknowledges the support by the National Natural Science Foundation of China(No. 11874011),
and C.W. is supported by the National Natural Science Foundation of China under Grant No. 11704093.


\appendix

\begin{figure*}[tbp]
\includegraphics[scale=0.5]{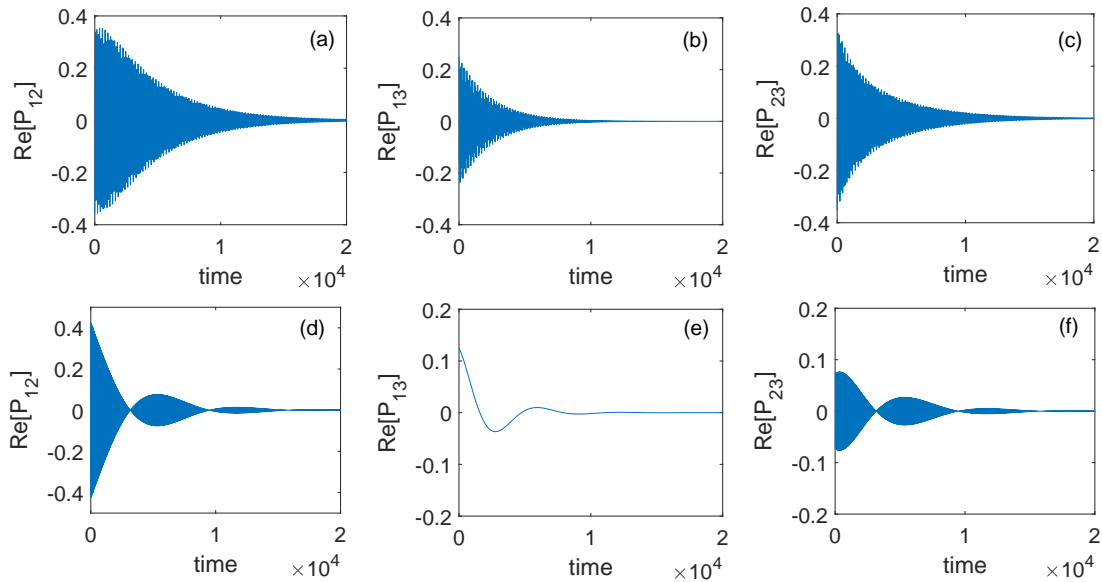}
\caption{(Color online)
 Dynamics of the off-diagonal elements of the collective qubits system $P_{ij}={\langle}\phi_i|\hat{\rho}_s(t{\rightarrow}\infty)|\phi_j{\rangle}$
 at resonance ($\varepsilon=0$) (a), (b) and (c), and at off-resonance ($\varepsilon=1$) (d), (e) and (f),
with the eigen-solution $\hat{H}_s|\phi_i{\rangle}=E_i|\phi_i{\rangle}$.
The initial state is given by $\hat{\rho}_s(0)=|1,-1{\rangle}{\langle}1,-1|$.
The other parameters are given by $N_s=2$,
$\Delta=1$, $\alpha=0.005$, $\omega_c=6$, $T_L=1.5$ and $T_R=0.5$.
}~\label{fig-app-a}
\end{figure*}

\section{Quantum thermal transfer within the Redfield scheme}

The nonequilibrium collective-qubit system is expressed as $\hat{H}=\hat{H}_s+\sum_{v=L,R}(\hat{H}^v_b+\hat{V}_v)$,
where the qubits Hamiltonian is given by
\begin{eqnarray}
\hat{H}_s={\varepsilon}\hat{J}_z+{\Delta}\hat{J}_x,
\end{eqnarray}
the $v$th thermal bath is $\hat{H}^v_b=\sum_k\omega_k\hat{b}^{\dag}_{k,v}\hat{b}_{k,v}$,
and the system-bath interaction is
\begin{eqnarray}~\label{app-a-vv1}
\hat{V}_v=2\hat{J}_z\sum_k(g_{kv}\hat{b}^{\dag}_{kv}+g^*_{kv}\hat{b}_{kv}).
\end{eqnarray}
To count the energy flow into the right bath including the full counting statistics, the total Hamiltonian is changed to
$\hat{H}(\chi)=e^{i\chi\hat{H}^R_b/2}\hat{H}e^{-i\chi\hat{H}^R_b/2}=\hat{H}_s+\sum_v(\hat{H}^v_b+\hat{V}^\chi_v)$,
where
\begin{eqnarray}~\label{app-a-vv1}
\hat{V}^\chi_v=2\hat{J}_z\sum_k(g_{kv}e^{i\frac{\chi\omega_k}{2}\delta_{v,R}}\hat{b}^{\dag}_{kv}
+H.c.).
\end{eqnarray}
Considering the weak qubit-bath interaction, we directly perturb the system-bath interaction at Eq.~(\ref{app-a-vv1}) up to the second order.
Then, based on the Born-Markov approximation, the Redfield equation is given by
\begin{eqnarray}
\frac{d\hat{\rho}_{\chi}(t)}{dt}&=&-i[\hat{H}_s,\hat{\rho}_{\chi}(t)]\\
&&-\sum_v\int^\infty_0{d\tau}\textrm{Tr}_b\{[\hat{V}^\chi_v,[\hat{V}^\chi_v(-\tau),\hat{\rho}_{\chi}(t){\otimes}\hat{\rho}_b]_\chi]_\chi\},\nonumber
\end{eqnarray}
with the commutating relation $[\hat{A}_\chi,\hat{B}_\chi]_\chi=\hat{A}_\chi\hat{B}_\chi-\hat{B}_\chi\hat{A}_{-\chi}$.
In the eigen-basis $\{|\phi_n{\rangle}\}$ with $\hat{H}_s|\phi_n{\rangle}=E_n|\phi_n{\rangle}$,
the dynamical equation of the system density matrix element is given by
\begin{eqnarray}
\frac{d\rho^{\chi}_{nn^{\prime}}}{dt}&=&-iE_{nn^{\prime}}\rho^{\chi}_{nn^{\prime}}
-\sum_{v,m,m^{\prime}}G_v(E_{mm^{\prime}})n_v(E_{mm^{\prime}})J^{nm}_{z}J^{mm^{\prime}}_{z}\rho^{\chi}_{m^{\prime}n^{\prime}}\\
&&-\sum_{v,m,m^{\prime}}G_v(E_{mm^{\prime}})(1+n_v(E_{mm^{\prime}}))J^{mm^{\prime}}_{z}J^{m^{\prime}n^{\prime}}_{z}\rho^{\chi}_{nm}\nonumber\\
&&+\sum_{v,m,m^{\prime}}G_v(E_{nm})n_v(E_{nm})e^{-iE_{nm}\chi\delta_{v,R}}J^{nm}_{z}J^{m^{\prime}n^{\prime}}_{z}\rho^{\chi}_{mm^{\prime}}\nonumber\\
&&+\sum_{v,m,m^{\prime}}G_v(E_{m^{\prime}n^{\prime}})(1+n_v(E_{m^{\prime}n^{\prime}}))e^{iE_{m^{\prime}n^{\prime}}\chi\delta_{v,R}}
J^{nm}_{z}J^{m^{\prime}n^{\prime}}_{z}\rho^{\chi}_{mm^{\prime}}\nonumber
\end{eqnarray}
where the counting-field dependent density matrix element is $\rho^{\chi}_{nn^{\prime}}={\langle}\phi_n|\hat{\rho}_{\chi}(t)|\phi_{n^{\prime}}{\rangle}$,
the transition rate is $G_v(\omega)=\pi\alpha_v\omega^3/\omega^2_c\exp{(-|\omega|/\omega_c)}$,
and the Bose-Einstein distribution function is $n_v(\omega)=1/[\exp(\omega/k_BT_v)-1]$.

At resonance ($\varepsilon=0$),
it is found that off-diagonal elements of the qubits system in eigen-space vanish at steady state in Figs.~\ref{fig-app-a}(a), (b) and (c),  in absence of the counting field ($\chi=0$).
Hence,  the steady state equation is given by
\begin{eqnarray}~\label{app-a-redfield}
&&\sum_{m{\neq}n,v}{G_v(E_{mn})}n_v(E_{mn})J^{nm}_{z}J^{mn}_{z}\rho_{nn}\nonumber\\
&&=\sum_{m{\neq}n,v}{G_v(E_{nm})}n_v(E_{nm})J^{nm}_{z}J^{mn}_{z}\rho_{mm}.
\end{eqnarray}
Actually, the system Hamiltonian at resonance is $\hat{H}_s=\Delta\hat{J}_x$, with the eigen-solution
$\hat{H_s}|j,m{\rangle}_x=\Delta{m}|j,m{\rangle}_x~(m=-N_s/2,-N_s/2+1,\cdots,N_s/2)$.
Thus, the coefficient $J^{nm}_z$ can be specified as  $J^{nm}_z=\frac{1}{2}(j^+_m\delta_{n,m+1}+j^-_m\delta_{n,m-1})$,
with $j^{\pm}_m=\sqrt{N_s/2(N_s/2+1)-m(m{\pm}1)}$.
Consequently, the steady state population can be analytically obtained as
\begin{eqnarray}
P^{ss}_m=x^{m+N_s/2}(1-x)/(1-x^{N_s+1}),
\end{eqnarray}
with the coefficient
\begin{eqnarray}
x=[\sum_{v=L,R}G_v(\Delta)n_v(\Delta)]/[\sum_{v=L,R}G_v(\Delta)(1+n_v(\Delta))].
\end{eqnarray}
Finally, the steady state heat flux is given by
\begin{eqnarray}~\label{app-a-jweak}
J_{\mathrm{weak}}=2(n_L(\Delta)-n_R(\Delta))\frac{G_L(\Delta)G_R(\Delta)}{G_L(\Delta)+G_R(\Delta)}I_N,
\end{eqnarray}
where the current factor is expressed as
\begin{eqnarray}
I_N=\frac{(N_s-\frac{2x}{1-x})/({1-x})^{N_s+1}+(\frac{x}{1-x})^{N_s+1}(N_s+\frac{2}{1-x})}
{(\frac{1}{1-x})^{N_s+1}-(\frac{x}{1-x})^{N_s+1}}.\nonumber\\
\end{eqnarray}

While at the off-resonant condition (e.g., $\varepsilon=1$), the off-diagonal elements after long time evolution also vanish,
shown at Figs.~\ref{fig-app-a}(d), (e) and (f).
Hence, the heat current into the right thermal bath is generally expressed as
\begin{eqnarray}
J_{weak}=2\sum_{n,m}E_{mn}\Gamma_R(E_{mn})(1+n_R(E_{mn}))J^{nm}_zJ^{mn}_zP^{ss}_m.\nonumber\\
\end{eqnarray}


\section{Quantum thermal transfer within the NIBA scheme}
Starting from the counting-field dependent Hamiltonian
$\hat{H}_\chi=\hat{H}_s+\sum_{u=L,R}(\hat{H}^u_b+\hat{V}_u(\chi))$ at Eq.~(\ref{hchi}),
we apply a canonical transformation $\hat{U}=\exp[i\hat{J}_z\sum_{u=L,R}\hat{B}_u(\chi)]$
with $\hat{B}_u(\chi)=2i\sum_k(\frac{g_{k,u}}{\omega_k}e^{i\frac{\omega_k\chi}{2}\delta_{u,R}}\hat{b}^{\dag}_{k,u}-H.c.)$,
to obtain the modified Hamiltonian of the whole system $\hat{H}^{\prime}_{\chi}=\hat{U}^{\dag}_\chi\hat{H}_{\chi}\hat{U}_\chi$ as
\begin{eqnarray}
\hat{H}^\prime_\chi=\hat{H}^{\textrm{NIBA}}_s+\sum_{u=L,R}(\hat{H}^u_b+\hat{V}^{\chi}_{sb}).
\end{eqnarray}
The modified system Hamiltonian is given by
\begin{eqnarray}
\hat{H}^{\textrm{NIBA}}_s=\varepsilon\hat{J}_z-{\xi}\hat{J}^2_z,
\end{eqnarray}
where the renormalization energy is $\xi=4\sum_{k,u=L,R}|g_{k,u}|^2/\omega_k$.
The eigen-solution is given by $\hat{H}^{\textrm{NIBA}}_s|\phi_n{\rangle}=E_n|\phi_n{\rangle}$
with $E_n=\varepsilon{n}-\xi{n}^2$ and $n=-N_s/2,{\cdots},N_s/2$.
The modified system-bath interaction is given by
\begin{eqnarray}
\hat{V}^{\chi}_{sb}=\frac{\Delta}{2}(e^{-i\hat{B}_{\chi}}\hat{J}_++e^{i\hat{B}_{\chi}}\hat{J}_-).
\end{eqnarray}
Hence, perturbing $\hat{V}^{\chi}_{sb}$  up to the second order, we obtain the quantum kinetic equation
\begin{eqnarray}~\label{qme-niba}
\frac{dP^\chi_n}{dt}=-(\kappa^-_{n-1}+\kappa^+_n)P^\chi_n+\kappa^-_{n}(\chi)P^\chi_{n+1}
+\kappa^+_{n-1}(\chi)P^\chi_{n-1},\nonumber\\
\end{eqnarray}
where the counting-field dependent population element is $P^\chi_n={\langle}\phi_n|\hat{\rho}_\chi(t)|\phi_n{\rangle}$
and the modified transition rates are
\begin{eqnarray}
\kappa^{\pm}_n(\chi)&=&\frac{(j^+_{n}\Delta)^2}{8\pi}\int^\infty_{-\infty}d\omega{e^{\mp{i\omega\chi}}}
C_L(\pm\omega\mp\Delta_n)C_R(\mp\omega),\nonumber\\
\end{eqnarray}
where the coefficient is $j^+_n={{N_s}/{2}({N_s}/{2}+1)-n(n{+}1)}$,
the energy gap $\Delta_n=E_{n+1}-E_n={\varepsilon}-\xi(2n+1)$,
the correlation function in the frequency domain is
\begin{eqnarray}
C_v(\omega)=\int^\infty_{-\infty}dte^{i\omega{t}}\eta^2_ve^{Q_v(t)},
\end{eqnarray}
with the renormalization factor
$\eta_v=\exp[-2\sum_k|\frac{g_{k,u}}{\omega_k}|^2(2n_{k,u}+1)]$,
and the correlation phase
\begin{eqnarray}
Q_{v}(\tau)=4\sum_{k}|\frac{g_{k,v}}{\omega_k}|^2[\cos\omega_k\tau(2n_{k,v}+1)-i\sin\omega_k\tau].\nonumber\\
\end{eqnarray}
In absence of the counting-field parameter ($\chi=0$), this modified kinetic equation is identical with the dynamical equation at Eq.~(\ref{niba1}).
Hence, the heat current at steady state is obtained as
\begin{eqnarray}
J&=&\sum^{\frac{N_s}{2}-1}_{n=-\frac{N_s}{2}}\frac{(j^+_n\Delta)^2}{8\pi}\int^\infty_{-\infty}
[C_L(-\omega+\Delta_{n})C_R(\omega)P^{ss}_{n+1}\nonumber\\
&&-C_L(\omega-\Delta_{n})C_R(-\omega)P^{ss}_{n}]{\omega}d\omega,
\end{eqnarray}
where the steady state population is given by $P^{ss}_{n}={\langle}N_s/2,n|\hat{\rho}_s(t{\rightarrow}\infty)|N_s/2,n{\rangle}$.

\end{document}